\documentclass[aps,prl,twocolumn,superscriptaddress,shopacs]{revtex4}
\usepackage{graphicx}
\usepackage{bm}
\usepackage{gensymb}
\usepackage[none]{hyphenat}
\usepackage{epstopdf}
\usepackage{multirow}
\usepackage{float}
\usepackage{tabularx}
\usepackage{gensymb}
\usepackage{color}
\usepackage{siunitx}
\usepackage{xr}
\usepackage{amsmath}
\begin{document}

\title{Robust block magnetism in the spin ladder compound BaFe$_2$Se$_3$ under hydrostatic  pressure}
\author{Shan Wu}
\email{shanwu@berkeley.edu}
\affiliation{Department of Physics, University of California, Berkeley, California, 94720, USA}

\author{Junjie Yin}
\affiliation{School of Physics, Sun Yat-Sen University, Guangzhou 510275, China}

\author{Thomas Smart}
\affiliation{Department of Earth and Planetary Science, University of California, Berkeley, California, 94720, USA}

\author{Arani Acharya}
\affiliation{Department of Physics, University of California, Berkeley, California, 94720, USA}

\author{Craig L Bull}
\author{Nicholas P Funnell}
\affiliation{ISIS Neutron and Muon Facility, Rutherford Appleton Laboratory, Chilton, Oxon, UK.}

\author{Thomas R Forrest}
\affiliation{Diamond Light Source, Chilton, Oxon, UK}

\author{Gediminas Simutis}
\author{Rustem Khasanov}
\affiliation{Laboratory for Muon Spin Spectroscopy, Paul Scherrer Institute, Villigen, PSI, Switzerland}

\author{Sylvia K. Lewin}
\affiliation{Department of Physics, University of California, Berkeley, California, 94720, USA}

\author{Meng Wang}
\affiliation{School of Physics, Sun Yat-Sen University, Guangzhou 510275, China}

\author{Benjamin A. Frandsen}
\affiliation{Department of Physics, Brigham Young University, Provo, Utah 84602, USA}

\author{Raymond Jeanloz}
\affiliation{Department of Earth and Planetary Science, University of California, Berkeley, California, 94720, USA}

\author{Robert J. Birgeneau}
\affiliation{Department of Physics, University of California, Berkeley, California, 94720, USA}

\date{\today}
\begin{abstract}
The majority of the iron-based superconductors (FeSCs) exhibit a two-dimensional square lattice structure. Recent reports of pressure-induced superconductivity in the spin-ladder system, BaFe$_2$X$_3$ (X =S,Se), introduce a quasi-one-dimensional prototype and an insulating parent compound to the  FeSCs. Here we report X-ray, neutron diffraction and muon spin relaxation experiments on BaFe$_2$Se$_3$ under hydrostatic pressure to investigate its magnetic and structural properties across the pressure-temperature phase diagram. A structural phase transition was identified at a pressure of 3.7(3) GPa. Neutron diffraction measurements at 6.8(3) GPa and 120 K show that the block magnetism persists even at these high pressures. A steady increase and then fast drop of the magnetic transition temperature $T\rm_N$ and greatly reduced moment above the pressure $P_s$ indicate potentially rich and competing phases close to the superconducting phase in this ladder system. 
\end{abstract}

\maketitle

In correlated electron materials, applied pressure or chemical substitution can alter the electronic structure and, concomitantly, the electron correlations, leading to wide varieties of electronic phases and phase transitions. These include  metal-insulator transitions, charge density wave order, antiferromagnetism, and superconductivity (SC) \cite{McWhanPRL,McWhanPRL2,VULETIC2006169,keimernature,stewartRMP,pcdRMP,dagottoRMP}. In iron-based superconductors, optimal superconductivity typically appears when the magnetic order is suppressed by the doping of carriers. Recent discoveries of pressure-induced SC around a critical pressure ($P_c$) of  10 GPa with the superconducting temperature $T\rm_c$ up to 24 K in BaFe$_2$X$_3$(X = S,Se) \cite{Takahashi2015,yamauchi2015,Ying2017} provide a new avenue for documenting the connection between magnetism and superconductivity without introducing any disorder by chemical doping. Importantly, these materials display a quasi-one-dimensional (1D) iron ladder structure \cite{Caron2011,Nambu2012} rather than the more usual square planar structure, and the parent compounds are insulators rather than poor metals. The reduced dimensionality and the metal-insulator transition preceding the superconducting phase resemble the characteristics of the cuprate system Sr$_{14-x}$Ca$_x$Cu$_{24}$O$_{41}$\cite{VULETIC2006169,nagata1998}. These ladder materials can thus provide important insights into the similarities and differences for both copper and iron-based superconductors. 

\begin{figure*}
\includegraphics[width=2.\columnwidth,clip,angle =0]{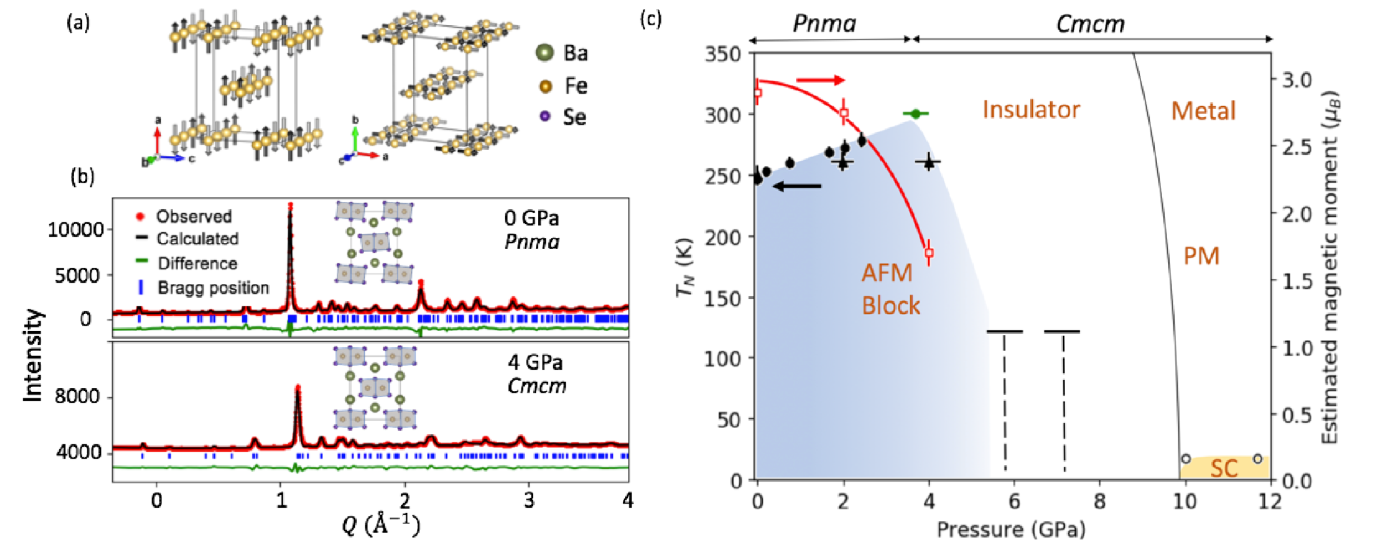}
\
\caption{\label{st_phase}  (a) Sketch of the block-type (left) and stripe-type (right) magnetic orders. (b) Room temperature X-ray diffraction patterns for BaFe$_2$Se$_3$ at $P$ = 0 and  4 GPa, fitted with $Pnma$ and $Cmcm$ models, respectively. Insets show corresponding structures viewed along the ladder direction. (c) Pressure versus temperature phase diagram constructed from the results in this paper and other work \cite{Ying2017}. The antiferromagnetic transition temperature $T\rm_N$ (black symbols, left axis) and estimated ordered moment (red symbols, right axis) are shown, with filled circles and triangles representing the $\mu SR$ and neutron diffraction results, respectively. The horizontal lines at the pressures of 5.5 and 6.8 GPa mark the lowest measured temperature of 120 K, at which we observed block-type short range magnetic correlations.  The green dot marks the structural transition ($P_s = 3.7(3)$ GPa) identified at 300 K.  The open circles denote $T\rm _c$ and black line marks the metal-insulator transition \cite{Ying2017}.  }
\label{phase}
\end{figure*}

Extensive experimental and theoretical work has been carried out on BaFe$_2$X$_3$ at ambient pressure \cite{Mourigal2015,Dong2014, Patel2018,Lovesey2016,Wang2017,konig2018,Aoyama2019,Herbrych2018,takubo2017}. In BaFe$_2$S$_3$, stripe-type antiferromagnetic (AFM) order  (Fig. \ref{st_phase} (a)) has been found below the N\'eel temperature ($T\rm_N$) of 119 K \cite{Takahashi2015,yamauchi2015}. In  contrast, BaFe$_2$Se$_3$ shows an exotic block-type magnetic order below 255 K \cite{Caron2011,Nambu2012}. The origin of this magnetic structure was ascribed theoretically to an orbital-selective Mott (OSMP) phase from multi-orbital Hubbard models for a 1D system \cite{luo2013,rincon2014,Herbrych2018}. 
Recent studies of the pressure-temperature ($P$-$T$) phase diagram by local magnetic probes have focused primarily on BaFe$_2$S$_3$, the first Fe-based spin-ladder system where pressure-induced superconductivity was reported \cite{chi2016,Zheng2018,Materne2019}. In BaFe$_2$Se$_3$, limited pressure-dependent experimental work has been carried out beyond transport measurements. The lack of information on the magnetic properties across the $P$-$T$ phase diagram is particularly notable. This is a result of the experimental challenges of performing neutron diffraction experiments at simultaneous high pressure ($>$ 2 GPa) and cryogenic temperatures, leaving a crucial gap in  experimental characterization of the iron-based spin ladder systems. In addition, the sensitivity of the magnetism to the local structure and stoichiometry in different samples may produce varied magnetic behaviors \cite{Zheng2018,saparov2011}. Thus, studies on the same sample are essential to establish a unified $P$-$T$ phase diagram.

In this Letter, we present a comprehensive characterization of the magneto-structural properties of BaFe$_2$Se$_3$ across a large region of the $P$-$T$ phase diagram using three complementary experimental probes at various pressure and temperature conditions: neutron powder diffraction (NPD) with pressures up to 6.8 GPa and temperatures down to 120 K, muon spin relaxation ($\mu SR$) measurements up to 2.43 GPa and down to 10 K, and X-ray powder diffraction (XRD) measurements up to 12 GPa at 300 K. Our measurements identify a structural transition from the $Pnma$ to $Cmcm$ space group at $P_s=$3.7(3) GPa. We observe a gradual enhancement of $T\rm_N$ with pressure up to $P_s$, followed by a considerable reduction of $T\rm_N$ above $P_s$. Intriguingly, the block-type magnetism in BaFe$_2$Se$_3$ remains stable up to the highest pressure measured by NPD (6.8 GPa), despite the fact that the crystallographic structure above $P_s$ is identical to that of BaFe$_2$S$_3$ with stripe-type magnetism. Comparison between these two ladder compounds yields important insights into the origin of the unusual block-type magnetism, and potentially the mechanism of superconductivity in these systems, greatly enriching the discussion of magnetism and superconductivity in iron-based materials.

A powder sample of BaFe$_2$Se$_3$ was synthesized by a self-flux solid state reaction \cite{Caron2011}. 
NPD measurements were conducted on the Pearl at the ISIS pulsed neutron facility in the UK \cite{bullpearl}. Diffraction data at ambient pressure were collected on HB3A, HFIR. $\mu SR$ measurements at ambient pressure and pressure-dependent were conducted on the beam line M20D at TRIUMF, Canada and the General Purpose Decay channel (GPD) instrument at the PSI, Switzerland \cite{khasanovgpd}, respectively.  Room temperature XRD experiments under pressure were performed at 12.2.2, Advanced Light Source. Experimental  details are described in the supplementary information (SI).

\begin{figure*}
\includegraphics[width=2\columnwidth,clip,angle =0]{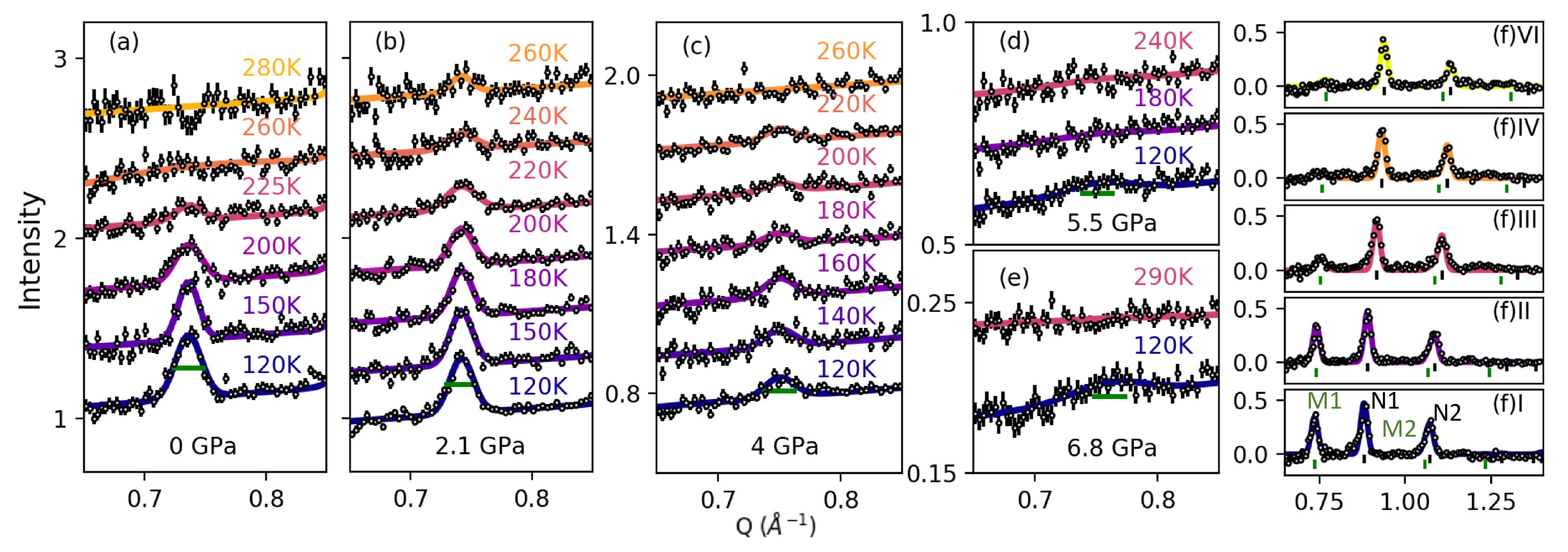}
\vspace{-1.5em}
\caption{\label{diff} Temperature dependence of neutron powder diffraction (NPD) data for $\rm BaFe_2Se_3$ at pressures of  (a) 0, (b) 2.1, (c) 4,  (d) 5.5, and (e) 6.8 GPa  collected at Pearl, ISIS \cite{bullpearl}. The additional peak at wave vector $Q = 0.75$ $ \rm \AA^{-1}$ that appears at low temperature can be indexed as ${Q}_{m1} $ = (0.5,0.5,0.5), consistent with the Fe$_4$ block spin structure. The data are normalized to 1 hour counting time and shifted a constant value vertically at each pressure for clarity. The solid curves are the results of fits using a single Gaussian peak. The horizontal bar represents the instrument resolution of 0.024 $\rm \AA^{-1}$ determined from a nearby nuclear Bragg peak. (f) Zoomed-out view of the NPD patterns at $T$ = 120 K for all pressures (0 to 6.8 GPa from the bottom to top). The colored curves are calculated patterns including both nuclear and magnetic phases. The peak positions with sufficient calculated/observed intensity are marked by vertical black (N1:(101), N2:(200)) and green lines (M1:($\frac{1}{2}\frac{1}{2}\frac{1}{2}$), M2:($\frac{3}{2}\frac{1}{2}\frac{1}{2}$)).} \label{diff}
\end{figure*}

The XRD (Fig. \ref{st_phase}(b)) and NPD  (Fig. S1) patterns at ambient conditions are well described by the $Pnma$ structure, as verified by Rietveld refinements using the FULLPROF Suite \cite{fullprof} with $\chi^2$ = 6.4. At $P$ = 4 GPa, the $Cmcm$ structure provides a significantly better fit ($\chi^2$ = 2.8) than $Pnma$ ($\chi^2$ = 6.6). The main difference between the two structures is that the ladder plane is tilted with respect to the crystallographic $a$ axis in the $Pnma$ structure (inset of Fig. \ref{st_phase} (b)). The critical pressure marking the transition from $Pnma$ to $Cmcm$ is estimated to be 3.7(3) GPa \cite{Svitlyk_2019}, based on inspection of the diffraction patterns (Fig. S2) and the corresponding refined lattice parameters versus pressure (see below).

In Figure \ref{diff}, we present NPD patterns at various pressures and temperatures. At ambient pressure, the diffraction patterns collected at and below 225 K show an additional Bragg reflection at $Q = 0.75$ $ \rm \AA^{-1}$, indexed as $Q_{m1}$ = ($\frac{1}{2}\frac{1}{2}\frac{1}{2}$). The calculated pattern is consistent with an Fe$_4$ block-type spin structure \cite{Caron2011, Caron2012,Nambu2012}, further confirmed by Rietveld refinement of the NPD data collected on HB3A at 1.5 K (Fig. S3) with a refined ordered moment of 2.9(3) $\mu_B$/Fe and transition temperature of 250 (5) K.
With increasing pressure, this magnetic peak (broad hump for  last two pressures) persists, indicating that the block magnetic state persists to the highest attainable pressure of 6.8 GPa.  Refined full width at half maximum (FWHM) values (Fig. S4) suggest that this peak is Q-resolution limited at 0, 2.1, 4 GPa (the horizontal bar in Fig. \ref{diff} (a-c)). However, at 5.5 and 6.8 GPa, the width of this magnetic peak is almost twice as wide as the instrumental $Q$-resolution of $0.025(2)\rm \AA^{-1}$(Fig. \ref{diff} (d-e)). Such broadening implies a crossover to short-range, block-type magnetic correlations, suggesting that any transition to true long-range magnetic order occurs below 120 K for 5.5 and 6.8 GPa.

At $P$ = 4 GPa ($> P_s$), the integrated intensity of ${Q}_{m1}$ is significantly reduced compared to that at lower pressures (33\%),  indicating a significant suppression of the ordered magnetic moment after the structural transition.  Assuming the block-type magnetic structure but varying the spin orientations, the fits to the data (Fig. \ref{diff} (f)III) and the difference data between low and high temperature (Fig. S5) for the first two magnetic peaks ($Q_{m1}$,$Q_{m2} = (\frac{3}{2}\frac{1}{2}\frac{1}{2})$) display slightly better results with spins perpendicular to the ladder direction ($a$ or $b$ axis) than along the ladder direction ($c$ axis). Importantly, above $P_s$, there is no clear observation of magnetic signal associated with the stripe structure found in BaFe$_2$S$_3$. Bragg reflections of the stripe structure would contribute significant intensity at $Q$ = 0.93 $\rm \AA^{-1}$and 1.31 $\rm \AA^{-1}$ (for $P$ = 6.8 GPa) for spins along the $c$ axis and $a$ or $b$ axes, respectively. Given statistics of the NPD data, we place an upper limit of 0.3 $\mu_B$ for any stripe-type moment in contrast to BaFe$_2$S$_3$ where stripe-type order is throughout the $Cmcm$ phase. This demonstrates that the structural symmetry is not solely responsible for the block-type order in BaFe$_2$Se$_3$.

As for the pressure-dependence of $T\rm_N$, the  $Q_{m1}$ peak at $T$ = 260 K is observed at 2.1 GPa but not 0 GPa (Fig. \ref{diff}(a-b)), indicating an increase in $T\rm_N$ with increasing pressure in the $Pnma$ phase. To verify it, we now turn to pressure-dependent $\mu SR$ measurements, which have finer control on the pressure and temperature. In Fig. \ref{musr}, we display temperature-dependent $\mu SR$ spectra from 0 up to 2.43 GPa collected in a zero field (ZF) configuration. The initial drop of the total asymmetry at low temperatures  originates from the long-range magnetic order; from this drop, we can obtain the magnetically ordered volume fraction under various pressures.

We modelled the $\mu SR$ spectra with an exponentially relaxing component and a Kubo-Toyabe component, corresponding to muons stopping in the sample and the sample holder, respectively \cite{khasanovgpd}. All fits were performed using MUSRFIT package \cite{suter2012}. We extracted the magnetic volume fraction ($f_{mag}$) as a function of temperature for pressures up to 2.43 GPa (Fig. \ref{dep} (b)). A bulk magnetic transition is manifest as a large increase in the $f_{mag}$ as the $T$ is lowered. With increasing pressure from 0 to 2.43 GPa, we observe the magnetic transition gradually moving to higher temperature. We quantify $T\rm _N$ at each pressure by defining it as the midpoint of the temperature region where the $f_{mag}$ changes (black star in Fig. \ref{dep} (b)), with error bars fixed at 20\% of the width of that temperature region. The results are shown as black dots in Fig. \ref{st_phase} (c).  We note that the $f_{mag}$ remains close to 1 at low temperature for all pressures, implying a fully ordered state below $T\rm_N$. 

\begin{figure}
\includegraphics[width=1.\columnwidth,clip,angle =0]{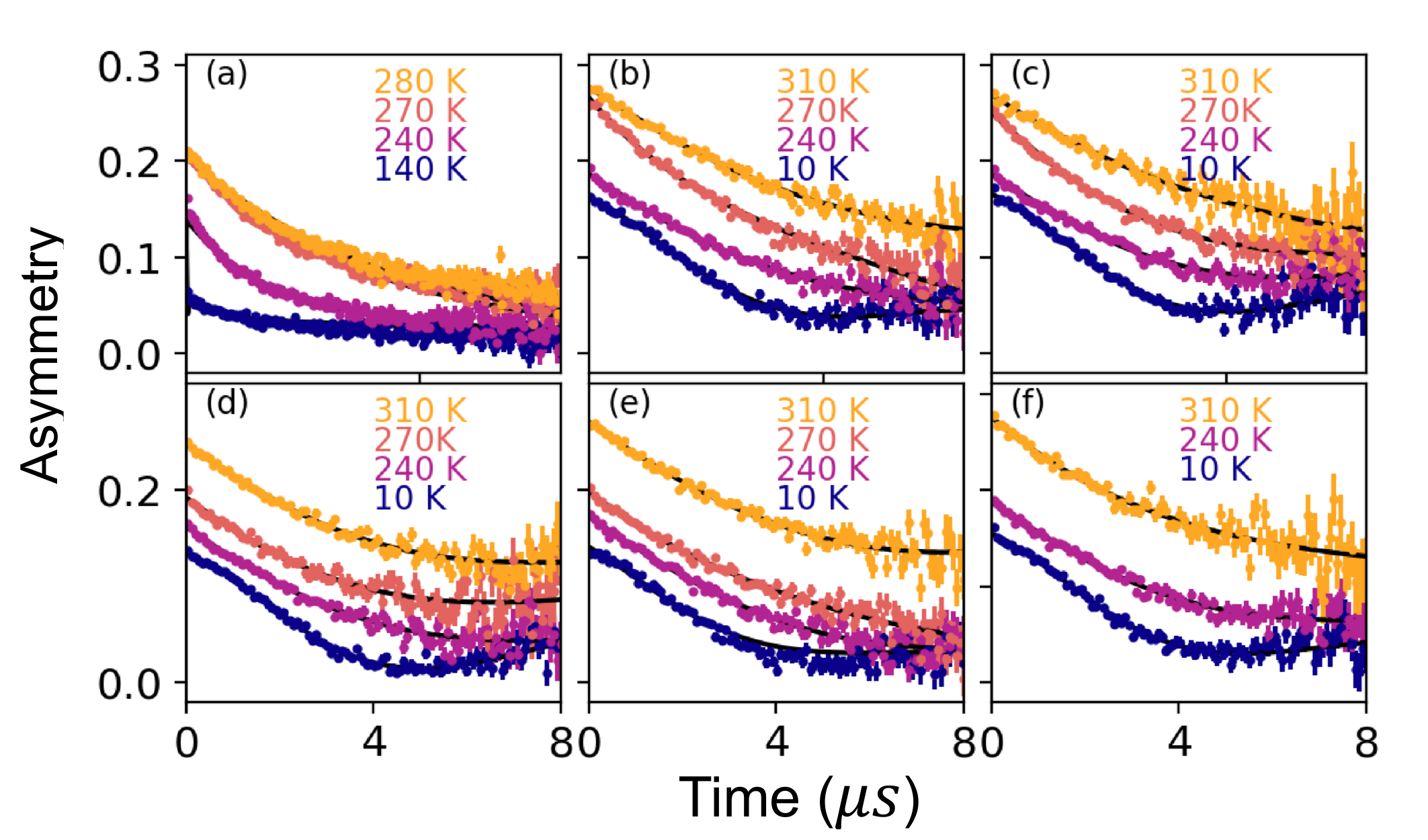}
\vspace{-1.5em}
\caption{\label{musr} Representative temperature-dependent  $\rm \mu SR$ spectra for $\rm BaFe_2Se_3$ at (a) ambient (data collected at TRIUMF) and measured pressures of (b) 0.19, (c) 0.76, (d) 1.67, (e) 2.03 , and (f) 2.43 GPa (data collected at PSI) in zero field configuration. The color dots and black lines represent data and the fits, respectively, at different temperatures.}
\label{musr}
\end{figure}

The smooth increase in $T\rm_N$ could be associated with local lattice changes driven by pressure. We fitted the Fe atomic coordinates $(x,y,z)$ in the $Pnma$ phase in the Rietveld refinements, in which the Wyckoff positions allow for a variation. These lead to the change of bond length between adjacent iron atoms along the ladder ($u$ and $v$) and the rung ($w$), with their ratios ($\frac{u}{w},\frac{v}{w}$) plotted in Fig. \ref{dep} (e). The increase of $\frac{u}{w}$ and decrease of $\frac{v}{w}$  with increasing pressure indicate a clear tendency to form the block state, which is also observed at ambient pressure with decreasing temperature \cite{Nambu2012}. Such a strong change due to the magnetoelastic coupling may account for the enhancement of $T\rm _N$. 

Having established the influence of pressure on $T\rm_N$, we now turn to the evolution of the ordered magnetic moment. We extracted the magnetic order parameter curves from the integrated area of $Q_{m1}$ (Fig. \ref{diff}) for measured pressures (Fig. \ref{dep} (a)). Fitting to a simple power law provides an estimated critical exponent of 0.29(2), close to the 3D  Ising value of 0.31. Comparison with the order parameter curve determined at ambient pressure, for which low-temperature data are available, allows us to estimate the low-temperature, saturated ordered moment at 2 and 4 GPa, as shown by red squares in Fig. \ref{st_phase} (c). The order parameter curves can be used to extract $T\rm_N$ as well; the results are plotted as black triangles in Fig. \ref{st_phase} (c) and are consistent with the $\mu SR$ results within the uncertainty. For 5.5 and 6.8 GPa, the symbols represent the upper boundary of the $T\rm_N$.

By combining the experimental observations from  XRD, NPD and $\mu SR$ measurements under hydrostatic pressure on the same sample, we can establish the $P$-$T$ phase diagram shown in Fig. \ref{st_phase} (c). Despite their different structures at ambient pressure, BaFe$_2$S$_3$ and BaFe$_2$Se$_3$ exhibit superconductivity at similar pressures and $T\rm _c$, as well as a common metal-insulator transition preceding the SC state. Another similarity is the initial enhancement of $T\rm_N$ \cite{Zheng2018,Materne2019} with increasing pressure, implying the magneto-elastic coupling inherent in the ladder system.
Several differences also exist between the two compounds. Along with its unusual block-type magnetism, BaFe$_2$Se$_3$ exhibits a potentially richer $P$-$T$ phase diagram. The structural phase transition from the $Pnma$ to the $Cmcm$ phase at $P_s = 3.7(3)$ GPa is confirmed by both X-ray and neutron diffraction measurements. No structural phase transition occurs in BaFe$_2$S$_3$, where the $Cmcm$ structure is present up to 12 GPa \cite{Kobayashi_2018}. In terms of the magnetic properties of BaFe$_2$Se$_3$, both the ordered moment $m$ and $T\rm _N$ drop quickly above $P_s$. 
This might associate with the Fe high-spin (HS) to low-spin (LS) transition induced by the structural transition, across which the structure has more compact stacking of the ladders (Fig. \ref{dep}(c)). The abrupt shortening is also observed in FeS and FeSe, where the structural transition is accompanied with a pressure-induced HS-LS transition \cite{Fei1892,rueff1999,lebert2018}. 
Despite the large reduction of $m$ and $T\rm _N$, the short-range block-type spin correlations remain present up to measured pressure of 6.8 GPa. Robust block magnetism in the $Cmcm$ phase, where the adjacent Fe-Fe distances are equivalent, implies that its origin may strongly relate to the electronic and orbital degrees of freedom. 

\begin{figure}
\includegraphics[width=1.\columnwidth,clip,angle =0]{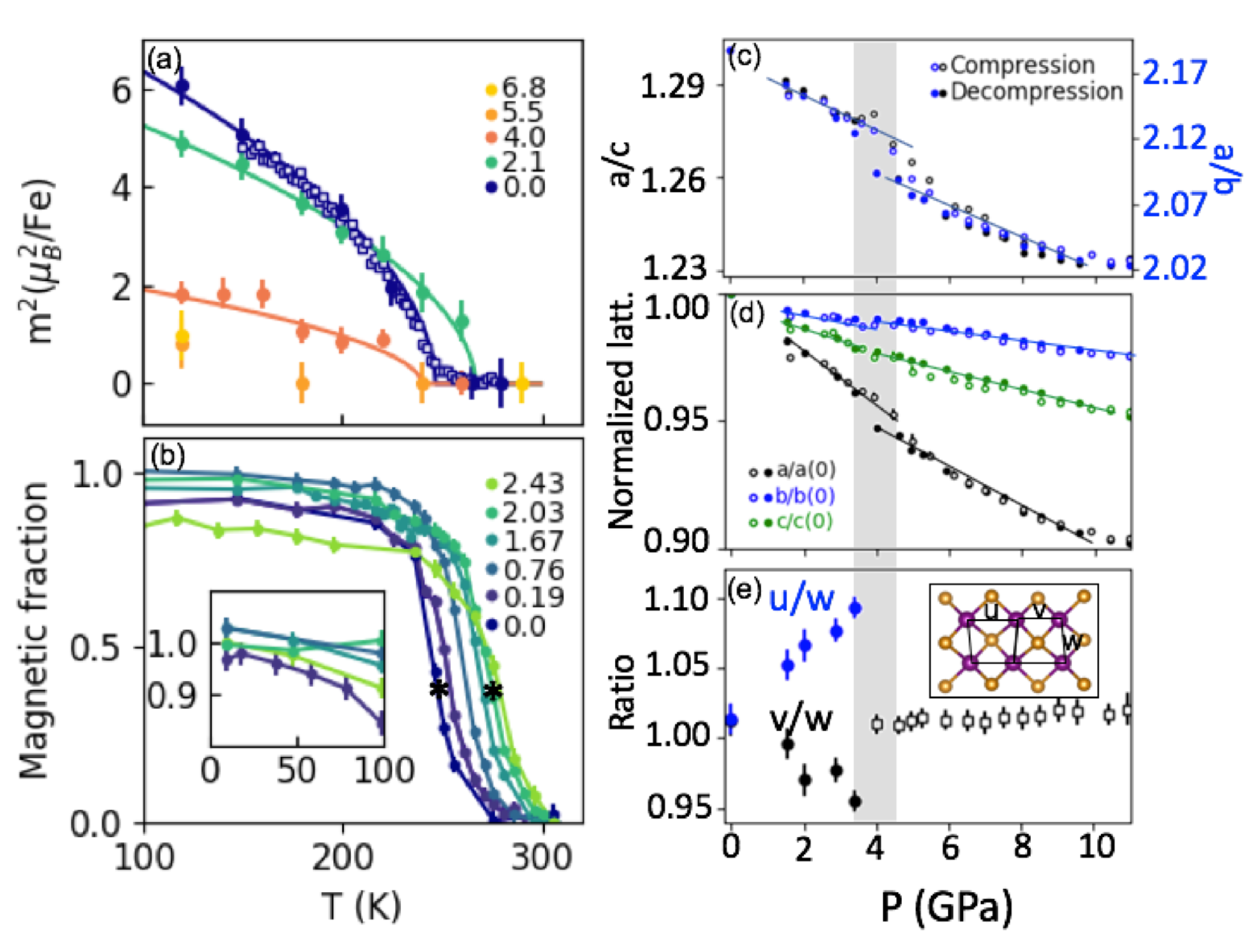} 
\caption{\label{dep} (a) Squared ordered magnetic moment ($ m^2$) versus temperature at applied  pressures marked in the units of GPa. The empty and filled symbols were neutron diffraction data collected at HB3A and Pearl, respectively. The solid lines are the power law fits of the form $m^2 \propto (T_c - T)^{2\beta}$. (b) Magnetically ordered volume fraction versus temperature determined from $\mu SR$ measurements at the indicated pressures. Inset shows the magnetic fraction below 100 K. (c) Ratios between out-of ladder ($a/c$) and within-ladder lattices ($a/b$). (d) Compressed (empty) and decompressed (filled) pressure-dependent lattice parameters normalized to the values at 0 GPa. The solid lines are the results of linear fits.
(e) Ratios of the Fe-Fe bond lengths along the ladder ($u$, $v$) and leg ($w$) direction versus decompressed pressure. Inset: the ladder plane in the $Pnma$ phase. The grey shadow marks the region of the structural transition.}
\label{dep}
\vspace{-1.em}
\end{figure}

Our results have vastly extended the range of pressures for which quantitative information exists on the structural phase transition, the evolution of the magnetic transition temperature and magnetic moment approaching the SC phase in the spin-ladder system BaFe$_2$Se$_3$. They have also made possible detailed comparisons with the sister compound BaFe$_2$S$_3$. The BaFe$_2$Se$_3$ $P$-$T$ phase diagram displays persistent block magnetism across a wide pressure range, characterized by an initial increase in $ T\rm_N$, followed by a strong reduction of $T\rm_N$ after the structural transition and a steadily reduced ordered moment with pressure. Such rich behavior is distinct from the dopant dependent phase diagram in other FeSCs with a 2D square lattice \cite{Fernandes2014}, where the magnetic ordered phase is continuously suppressed close to the SC phase. This may be associated with enhanced electron correlations in the quasi-1D ladder approaching the insulator-metal transition.  The next frontier requires extending measurements into the SC phase starting from 10 GPa \cite{Ying2017}. Whether the driving force of the superconductivity is induced by magnetic fluctuations due to the competition between block and stripe magnetic states proposed theoretically \cite{Zhang2018} or other electronic phases is still an open question. 

The authors would like to acknowledge the beam line support from Martin Kunz and Andrew Doran at ALS, Gerald Morris at TRIUMF and Keith Taddei at HFIR. The authors would like to also thank STFC for provision of beam time at the ISIS Neutron and Muon Facility\cite{isisnm}.This work is funded by the U.S. Department of Energy, Office of Science, Office of Basic Energy Sciences, Materials Sciences and Engineering Division under Contract No. DE-AC02-05-CH11231 within the Quantum Materials Program (KC2202). M. W. and J. J. Y were supported by NSFC-11904414 and NSF of Guangdong under Contract No. 2018A030313055. S.K.L. was supported by an AAUW Dissertation Fellowship.

\bibliography{bibfile}

\begin{thebibliography}{41}
\expandafter\ifx\csname natexlab\endcsname\relax\def\natexlab#1{#1}\fi
\expandafter\ifx\csname bibnamefont\endcsname\relax
  \def\bibnamefont#1{#1}\fi
\expandafter\ifx\csname bibfnamefont\endcsname\relax
  \def\bibfnamefont#1{#1}\fi
\expandafter\ifx\csname citenamefont\endcsname\relax
  \def\citenamefont#1{#1}\fi
\expandafter\ifx\csname url\endcsname\relax
  \def\url#1{\texttt{#1}}\fi
\expandafter\ifx\csname urlprefix\endcsname\relax\def\urlprefix{URL }\fi
\providecommand{\bibinfo}[2]{#2}
\providecommand{\eprint}[2][]{\url{#2}}

\bibitem[{\citenamefont{McWhan et~al.}(1971)\citenamefont{McWhan, Remeika,
  Rice, Brinkman, Maita, and Menth}}]{McWhanPRL}
\bibinfo{author}{\bibfnamefont{D.~B.} \bibnamefont{McWhan}},
  \bibinfo{author}{\bibfnamefont{J.~P.} \bibnamefont{Remeika}},
  \bibinfo{author}{\bibfnamefont{T.~M.} \bibnamefont{Rice}},
  \bibinfo{author}{\bibfnamefont{W.~F.} \bibnamefont{Brinkman}},
  \bibinfo{author}{\bibfnamefont{J.~P.} \bibnamefont{Maita}}, \bibnamefont{and}
  \bibinfo{author}{\bibfnamefont{A.}~\bibnamefont{Menth}},
  \bibinfo{journal}{Phys. Rev. Lett.} \textbf{\bibinfo{volume}{27}},
  \bibinfo{pages}{941} (\bibinfo{year}{1971}).

\bibitem[{\citenamefont{McWhan et~al.}(1973)\citenamefont{McWhan, Menth,
  Remeika, Brinkman, and Rice}}]{McWhanPRL2}
\bibinfo{author}{\bibfnamefont{D.~B.} \bibnamefont{McWhan}},
  \bibinfo{author}{\bibfnamefont{A.}~\bibnamefont{Menth}},
  \bibinfo{author}{\bibfnamefont{J.~P.} \bibnamefont{Remeika}},
  \bibinfo{author}{\bibfnamefont{W.~F.} \bibnamefont{Brinkman}},
  \bibnamefont{and} \bibinfo{author}{\bibfnamefont{T.~M.} \bibnamefont{Rice}},
  \bibinfo{journal}{Phys. Rev. B} \textbf{\bibinfo{volume}{7}},
  \bibinfo{pages}{1920} (\bibinfo{year}{1973}).

\bibitem[{VUL(2006)}]{VULETIC2006169}
\bibinfo{journal}{Physics Reports} \textbf{\bibinfo{volume}{428}},
  \bibinfo{pages}{169 } (\bibinfo{year}{2006}), ISSN \bibinfo{issn}{0370-1573}.

\bibitem[{\citenamefont{Keimer et~al.}(2015)\citenamefont{Keimer, Kivelson,
  Norman, Uchida, and Zaanen}}]{keimernature}
\bibinfo{author}{\bibfnamefont{B.}~\bibnamefont{Keimer}},
  \bibinfo{author}{\bibfnamefont{S.}~\bibnamefont{Kivelson}},
  \bibinfo{author}{\bibfnamefont{M.}~\bibnamefont{Norman}},
  \bibinfo{author}{\bibfnamefont{S.}~\bibnamefont{Uchida}}, \bibnamefont{and}
  \bibinfo{author}{\bibfnamefont{J.}~\bibnamefont{Zaanen}},
  \bibinfo{journal}{Nature} \textbf{\bibinfo{volume}{518}}, \bibinfo{pages}{179
  EP} (\bibinfo{year}{2015}).

\bibitem[{\citenamefont{Stewart}(2011)}]{stewartRMP}
\bibinfo{author}{\bibfnamefont{G.~R.} \bibnamefont{Stewart}},
  \bibinfo{journal}{Rev. Mod. Phys.} \textbf{\bibinfo{volume}{83}},
  \bibinfo{pages}{1589} (\bibinfo{year}{2011}).

\bibitem[{\citenamefont{Dai}(2015)}]{pcdRMP}
\bibinfo{author}{\bibfnamefont{P.}~\bibnamefont{Dai}}, \bibinfo{journal}{Rev.
  Mod. Phys.} \textbf{\bibinfo{volume}{87}}, \bibinfo{pages}{855}
  (\bibinfo{year}{2015}).

\bibitem[{\citenamefont{Dagotto}(2013)}]{dagottoRMP}
\bibinfo{author}{\bibfnamefont{E.}~\bibnamefont{Dagotto}},
  \bibinfo{journal}{Rev. Mod. Phys.} \textbf{\bibinfo{volume}{85}},
  \bibinfo{pages}{849} (\bibinfo{year}{2013}).

\bibitem[{\citenamefont{Takahashi et~al.}(2015)\citenamefont{Takahashi,
  Sugimoto, Nambu, Yamauchi, Hirata, Kawakami, Avdeev, Matsubayashi, Du,
  Kawashima et~al.}}]{Takahashi2015}
\bibinfo{author}{\bibfnamefont{H.}~\bibnamefont{Takahashi}},
  \bibinfo{author}{\bibfnamefont{A.}~\bibnamefont{Sugimoto}},
  \bibinfo{author}{\bibfnamefont{Y.}~\bibnamefont{Nambu}},
  \bibinfo{author}{\bibfnamefont{T.}~\bibnamefont{Yamauchi}},
  \bibinfo{author}{\bibfnamefont{Y.}~\bibnamefont{Hirata}},
  \bibinfo{author}{\bibfnamefont{T.}~\bibnamefont{Kawakami}},
  \bibinfo{author}{\bibfnamefont{M.}~\bibnamefont{Avdeev}},
  \bibinfo{author}{\bibfnamefont{K.}~\bibnamefont{Matsubayashi}},
  \bibinfo{author}{\bibfnamefont{F.}~\bibnamefont{Du}},
  \bibinfo{author}{\bibfnamefont{C.}~\bibnamefont{Kawashima}},
  \bibnamefont{et~al.}, \bibinfo{journal}{Nature Materials}
  \textbf{\bibinfo{volume}{14}}, \bibinfo{pages}{1008} (\bibinfo{year}{2015}),
  ISSN \bibinfo{issn}{1476-1122}.

\bibitem[{\citenamefont{Yamauchi et~al.}(2015)\citenamefont{Yamauchi, Hirata,
  Ueda, and Ohgushi}}]{yamauchi2015}
\bibinfo{author}{\bibfnamefont{T.}~\bibnamefont{Yamauchi}},
  \bibinfo{author}{\bibfnamefont{Y.}~\bibnamefont{Hirata}},
  \bibinfo{author}{\bibfnamefont{Y.}~\bibnamefont{Ueda}}, \bibnamefont{and}
  \bibinfo{author}{\bibfnamefont{K.}~\bibnamefont{Ohgushi}},
  \bibinfo{journal}{Phys. Rev. Lett.} \textbf{\bibinfo{volume}{115}},
  \bibinfo{pages}{246402} (\bibinfo{year}{2015}).

\bibitem[{\citenamefont{Ying et~al.}(2017)\citenamefont{Ying, Lei, Petrovic,
  Xiao, and Struzhkin}}]{Ying2017}
\bibinfo{author}{\bibfnamefont{J.}~\bibnamefont{Ying}},
  \bibinfo{author}{\bibfnamefont{H.}~\bibnamefont{Lei}},
  \bibinfo{author}{\bibfnamefont{C.}~\bibnamefont{Petrovic}},
  \bibinfo{author}{\bibfnamefont{Y.}~\bibnamefont{Xiao}}, \bibnamefont{and}
  \bibinfo{author}{\bibfnamefont{V.~V.} \bibnamefont{Struzhkin}},
  \bibinfo{journal}{Physical Review B Rapid Communication}
  \textbf{\bibinfo{volume}{95}}, \bibinfo{pages}{241109(R)}
  (\bibinfo{year}{2017}).

\bibitem[{\citenamefont{Caron et~al.}(2011)\citenamefont{Caron, Neilson,
  Miller, Llobet, and McQueen}}]{Caron2011}
\bibinfo{author}{\bibfnamefont{J.~M.} \bibnamefont{Caron}},
  \bibinfo{author}{\bibfnamefont{J.~R.} \bibnamefont{Neilson}},
  \bibinfo{author}{\bibfnamefont{D.~C.} \bibnamefont{Miller}},
  \bibinfo{author}{\bibfnamefont{A.}~\bibnamefont{Llobet}}, \bibnamefont{and}
  \bibinfo{author}{\bibfnamefont{T.~M.} \bibnamefont{McQueen}},
  \bibinfo{journal}{Phys. Rev. B} \textbf{\bibinfo{volume}{84}},
  \bibinfo{pages}{180409(R)} (\bibinfo{year}{2011}).

\bibitem[{\citenamefont{Nambu et~al.}(2012)\citenamefont{Nambu, Ohgushi,
  Suzuki, Du, Avdeev, Uwatoko, Munakata, Fukazawa, Chi, Ueda
  et~al.}}]{Nambu2012}
\bibinfo{author}{\bibfnamefont{Y.}~\bibnamefont{Nambu}},
  \bibinfo{author}{\bibfnamefont{K.}~\bibnamefont{Ohgushi}},
  \bibinfo{author}{\bibfnamefont{S.}~\bibnamefont{Suzuki}},
  \bibinfo{author}{\bibfnamefont{F.}~\bibnamefont{Du}},
  \bibinfo{author}{\bibfnamefont{M.}~\bibnamefont{Avdeev}},
  \bibinfo{author}{\bibfnamefont{Y.}~\bibnamefont{Uwatoko}},
  \bibinfo{author}{\bibfnamefont{K.}~\bibnamefont{Munakata}},
  \bibinfo{author}{\bibfnamefont{H.}~\bibnamefont{Fukazawa}},
  \bibinfo{author}{\bibfnamefont{S.}~\bibnamefont{Chi}},
  \bibinfo{author}{\bibfnamefont{Y.}~\bibnamefont{Ueda}}, \bibnamefont{et~al.},
  \bibinfo{journal}{Phys. Rev. B} \textbf{\bibinfo{volume}{85}},
  \bibinfo{pages}{064413} (\bibinfo{year}{2012}).

\bibitem[{\citenamefont{Nagata et~al.}(1998)\citenamefont{Nagata, Uehara, Goto,
  Akimitsu, Motoyama, Eisaki, Uchida, Takahashi, Nakanishi, and
  M\^ori}}]{nagata1998}
\bibinfo{author}{\bibfnamefont{T.}~\bibnamefont{Nagata}},
  \bibinfo{author}{\bibfnamefont{M.}~\bibnamefont{Uehara}},
  \bibinfo{author}{\bibfnamefont{J.}~\bibnamefont{Goto}},
  \bibinfo{author}{\bibfnamefont{J.}~\bibnamefont{Akimitsu}},
  \bibinfo{author}{\bibfnamefont{N.}~\bibnamefont{Motoyama}},
  \bibinfo{author}{\bibfnamefont{H.}~\bibnamefont{Eisaki}},
  \bibinfo{author}{\bibfnamefont{S.}~\bibnamefont{Uchida}},
  \bibinfo{author}{\bibfnamefont{H.}~\bibnamefont{Takahashi}},
  \bibinfo{author}{\bibfnamefont{T.}~\bibnamefont{Nakanishi}},
  \bibnamefont{and} \bibinfo{author}{\bibfnamefont{N.}~\bibnamefont{M\^ori}},
  \bibinfo{journal}{Phys. Rev. Lett.} \textbf{\bibinfo{volume}{81}},
  \bibinfo{pages}{1090} (\bibinfo{year}{1998}).

\bibitem[{\citenamefont{Mourigal et~al.}(2015)\citenamefont{Mourigal, Wu,
  Stone, Neilson, Caron, McQueen, and Broholm}}]{Mourigal2015}
\bibinfo{author}{\bibfnamefont{M.}~\bibnamefont{Mourigal}},
  \bibinfo{author}{\bibfnamefont{S.}~\bibnamefont{Wu}},
  \bibinfo{author}{\bibfnamefont{M.~B.} \bibnamefont{Stone}},
  \bibinfo{author}{\bibfnamefont{J.~R.} \bibnamefont{Neilson}},
  \bibinfo{author}{\bibfnamefont{J.~M.} \bibnamefont{Caron}},
  \bibinfo{author}{\bibfnamefont{T.~M.} \bibnamefont{McQueen}},
  \bibnamefont{and} \bibinfo{author}{\bibfnamefont{C.~L.}
  \bibnamefont{Broholm}}, \bibinfo{journal}{Phys. Rev. Lett.}
  \textbf{\bibinfo{volume}{115}}, \bibinfo{pages}{047401}
  (\bibinfo{year}{2015}).

\bibitem[{\citenamefont{Dong et~al.}(2014)\citenamefont{Dong, Liu, and
  Dagotto}}]{Dong2014}
\bibinfo{author}{\bibfnamefont{S.}~\bibnamefont{Dong}},
  \bibinfo{author}{\bibfnamefont{J.-M.} \bibnamefont{Liu}}, \bibnamefont{and}
  \bibinfo{author}{\bibfnamefont{E.}~\bibnamefont{Dagotto}},
  \bibinfo{journal}{Phys. Rev. Lett.} \textbf{\bibinfo{volume}{113}},
  \bibinfo{pages}{187204} (\bibinfo{year}{2014}).

\bibitem[{\citenamefont{Patel et~al.}(2018)\citenamefont{Patel, Nocera,
  Alvarez, Moreo, Johnston, and Dagotto}}]{Patel2018}
\bibinfo{author}{\bibfnamefont{N.~D.} \bibnamefont{Patel}},
  \bibinfo{author}{\bibfnamefont{A.}~\bibnamefont{Nocera}},
  \bibinfo{author}{\bibfnamefont{G.}~\bibnamefont{Alvarez}},
  \bibinfo{author}{\bibfnamefont{A.}~\bibnamefont{Moreo}},
  \bibinfo{author}{\bibfnamefont{S.}~\bibnamefont{Johnston}}, \bibnamefont{and}
  \bibinfo{author}{\bibfnamefont{E.}~\bibnamefont{Dagotto}},
  \bibinfo{type}{Tech. Rep.} (\bibinfo{year}{2018}), \eprint{1807.10419v2}.

\bibitem[{\citenamefont{Lovesey et~al.}(2016)\citenamefont{Lovesey, Khalyavin,
  and van~der Laan}}]{Lovesey2016}
\bibinfo{author}{\bibfnamefont{S.~W.} \bibnamefont{Lovesey}},
  \bibinfo{author}{\bibfnamefont{D.~D.} \bibnamefont{Khalyavin}},
  \bibnamefont{and} \bibinfo{author}{\bibfnamefont{G.}~\bibnamefont{van~der
  Laan}}, \bibinfo{journal}{Physica Scripta} \textbf{\bibinfo{volume}{91}},
  \bibinfo{pages}{015803} (\bibinfo{year}{2016}), ISSN
  \bibinfo{issn}{0031-8949}.

\bibitem[{\citenamefont{Wang et~al.}(2017)\citenamefont{Wang, Jin, Yi, Song,
  Jiang, Zhang, Sun, Luo, Christianson, Bourret-Courchesne et~al.}}]{Wang2017}
\bibinfo{author}{\bibfnamefont{M.}~\bibnamefont{Wang}},
  \bibinfo{author}{\bibfnamefont{S.~J.} \bibnamefont{Jin}},
  \bibinfo{author}{\bibfnamefont{M.}~\bibnamefont{Yi}},
  \bibinfo{author}{\bibfnamefont{Y.}~\bibnamefont{Song}},
  \bibinfo{author}{\bibfnamefont{H.~C.} \bibnamefont{Jiang}},
  \bibinfo{author}{\bibfnamefont{W.~L.} \bibnamefont{Zhang}},
  \bibinfo{author}{\bibfnamefont{H.~L.} \bibnamefont{Sun}},
  \bibinfo{author}{\bibfnamefont{H.~Q.} \bibnamefont{Luo}},
  \bibinfo{author}{\bibfnamefont{A.~D.} \bibnamefont{Christianson}},
  \bibinfo{author}{\bibfnamefont{E.}~\bibnamefont{Bourret-Courchesne}},
  \bibnamefont{et~al.}, \bibinfo{journal}{Phys. Rev. B}
  \textbf{\bibinfo{volume}{95}}, \bibinfo{pages}{060502(R)}
  (\bibinfo{year}{2017}).

\bibitem[{\citenamefont{K\"onig et~al.}(2018)\citenamefont{K\"onig, Tsvelik,
  and Coleman}}]{konig2018}
\bibinfo{author}{\bibfnamefont{E.~J.} \bibnamefont{K\"onig}},
  \bibinfo{author}{\bibfnamefont{A.~M.} \bibnamefont{Tsvelik}},
  \bibnamefont{and} \bibinfo{author}{\bibfnamefont{P.}~\bibnamefont{Coleman}},
  \bibinfo{journal}{Phys. Rev. B} \textbf{\bibinfo{volume}{98}},
  \bibinfo{pages}{184517} (\bibinfo{year}{2018}).

\bibitem[{\citenamefont{Aoyama et~al.}(2019)\citenamefont{Aoyama, Imaizumi,
  Togashi, Sato, Hashizume, Nambu, Hirata, Matsubara, and
  Ohgushi}}]{Aoyama2019}
\bibinfo{author}{\bibfnamefont{T.}~\bibnamefont{Aoyama}},
  \bibinfo{author}{\bibfnamefont{S.}~\bibnamefont{Imaizumi}},
  \bibinfo{author}{\bibfnamefont{T.}~\bibnamefont{Togashi}},
  \bibinfo{author}{\bibfnamefont{Y.}~\bibnamefont{Sato}},
  \bibinfo{author}{\bibfnamefont{K.}~\bibnamefont{Hashizume}},
  \bibinfo{author}{\bibfnamefont{Y.}~\bibnamefont{Nambu}},
  \bibinfo{author}{\bibfnamefont{Y.}~\bibnamefont{Hirata}},
  \bibinfo{author}{\bibfnamefont{M.}~\bibnamefont{Matsubara}},
  \bibnamefont{and} \bibinfo{author}{\bibfnamefont{K.}~\bibnamefont{Ohgushi}},
  \bibinfo{type}{Tech. Rep.} (\bibinfo{year}{2019}), \eprint{1902.10900v1}.

\bibitem[{\citenamefont{Herbrych et~al.}(2018)\citenamefont{Herbrych, Kaushal,
  Nocera, Alvarez, Moreo, and Dagotto}}]{Herbrych2018}
\bibinfo{author}{\bibfnamefont{J.}~\bibnamefont{Herbrych}},
  \bibinfo{author}{\bibfnamefont{N.}~\bibnamefont{Kaushal}},
  \bibinfo{author}{\bibfnamefont{A.}~\bibnamefont{Nocera}},
  \bibinfo{author}{\bibfnamefont{G.}~\bibnamefont{Alvarez}},
  \bibinfo{author}{\bibfnamefont{A.}~\bibnamefont{Moreo}}, \bibnamefont{and}
  \bibinfo{author}{\bibfnamefont{E.}~\bibnamefont{Dagotto}},
  \bibinfo{journal}{Nature Communications} p. \bibinfo{pages}{3736}
  (\bibinfo{year}{2018}), ISSN \bibinfo{issn}{2041-1723}.

\bibitem[{\citenamefont{Takubo et~al.}(2017)\citenamefont{Takubo, Yokoyama,
  Wadati, Iwasaki, Mizokawa, Boyko, Sutarto, He, Hashizume, Imaizumi
  et~al.}}]{takubo2017}
\bibinfo{author}{\bibfnamefont{K.}~\bibnamefont{Takubo}},
  \bibinfo{author}{\bibfnamefont{Y.}~\bibnamefont{Yokoyama}},
  \bibinfo{author}{\bibfnamefont{H.}~\bibnamefont{Wadati}},
  \bibinfo{author}{\bibfnamefont{S.}~\bibnamefont{Iwasaki}},
  \bibinfo{author}{\bibfnamefont{T.}~\bibnamefont{Mizokawa}},
  \bibinfo{author}{\bibfnamefont{T.}~\bibnamefont{Boyko}},
  \bibinfo{author}{\bibfnamefont{R.}~\bibnamefont{Sutarto}},
  \bibinfo{author}{\bibfnamefont{F.}~\bibnamefont{He}},
  \bibinfo{author}{\bibfnamefont{K.}~\bibnamefont{Hashizume}},
  \bibinfo{author}{\bibfnamefont{S.}~\bibnamefont{Imaizumi}},
  \bibnamefont{et~al.}, \bibinfo{journal}{Phys. Rev. B}
  \textbf{\bibinfo{volume}{96}}, \bibinfo{pages}{115157}
  (\bibinfo{year}{2017}).

\bibitem[{\citenamefont{Luo et~al.}(2013)\citenamefont{Luo, Nicholson,
  Rinc\'on, Liang, Riera, Alvarez, Wang, Ku, Samolyuk, Moreo et~al.}}]{luo2013}
\bibinfo{author}{\bibfnamefont{Q.}~\bibnamefont{Luo}},
  \bibinfo{author}{\bibfnamefont{A.}~\bibnamefont{Nicholson}},
  \bibinfo{author}{\bibfnamefont{J.}~\bibnamefont{Rinc\'on}},
  \bibinfo{author}{\bibfnamefont{S.}~\bibnamefont{Liang}},
  \bibinfo{author}{\bibfnamefont{J.}~\bibnamefont{Riera}},
  \bibinfo{author}{\bibfnamefont{G.}~\bibnamefont{Alvarez}},
  \bibinfo{author}{\bibfnamefont{L.}~\bibnamefont{Wang}},
  \bibinfo{author}{\bibfnamefont{W.}~\bibnamefont{Ku}},
  \bibinfo{author}{\bibfnamefont{G.~D.} \bibnamefont{Samolyuk}},
  \bibinfo{author}{\bibfnamefont{A.}~\bibnamefont{Moreo}},
  \bibnamefont{et~al.}, \bibinfo{journal}{Phys. Rev. B}
  \textbf{\bibinfo{volume}{87}}, \bibinfo{pages}{024404}
  (\bibinfo{year}{2013}).

\bibitem[{\citenamefont{Rinc\'on et~al.}(2014)\citenamefont{Rinc\'on, Moreo,
  Alvarez, and Dagotto}}]{rincon2014}
\bibinfo{author}{\bibfnamefont{J.}~\bibnamefont{Rinc\'on}},
  \bibinfo{author}{\bibfnamefont{A.}~\bibnamefont{Moreo}},
  \bibinfo{author}{\bibfnamefont{G.}~\bibnamefont{Alvarez}}, \bibnamefont{and}
  \bibinfo{author}{\bibfnamefont{E.}~\bibnamefont{Dagotto}},
  \bibinfo{journal}{Phys. Rev. Lett.} \textbf{\bibinfo{volume}{112}},
  \bibinfo{pages}{106405} (\bibinfo{year}{2014}).

\bibitem[{\citenamefont{Chi et~al.}(2016)\citenamefont{Chi, Uwatoko, Cao,
  Hirata, Hashizume, Aoyama, and Ohgushi}}]{chi2016}
\bibinfo{author}{\bibfnamefont{S.}~\bibnamefont{Chi}},
  \bibinfo{author}{\bibfnamefont{Y.}~\bibnamefont{Uwatoko}},
  \bibinfo{author}{\bibfnamefont{H.}~\bibnamefont{Cao}},
  \bibinfo{author}{\bibfnamefont{Y.}~\bibnamefont{Hirata}},
  \bibinfo{author}{\bibfnamefont{K.}~\bibnamefont{Hashizume}},
  \bibinfo{author}{\bibfnamefont{T.}~\bibnamefont{Aoyama}}, \bibnamefont{and}
  \bibinfo{author}{\bibfnamefont{K.}~\bibnamefont{Ohgushi}},
  \bibinfo{journal}{Phys. Rev. Lett.} \textbf{\bibinfo{volume}{117}},
  \bibinfo{pages}{047003} (\bibinfo{year}{2016}).

\bibitem[{\citenamefont{Zheng et~al.}(2018)\citenamefont{Zheng, Frandsen, Wu,
  Yi, Wu, Huang, Bourret-Courchesne, Simutis, Khasanov, Yao
  et~al.}}]{Zheng2018}
\bibinfo{author}{\bibfnamefont{L.}~\bibnamefont{Zheng}},
  \bibinfo{author}{\bibfnamefont{B.~A.} \bibnamefont{Frandsen}},
  \bibinfo{author}{\bibfnamefont{C.}~\bibnamefont{Wu}},
  \bibinfo{author}{\bibfnamefont{M.}~\bibnamefont{Yi}},
  \bibinfo{author}{\bibfnamefont{S.}~\bibnamefont{Wu}},
  \bibinfo{author}{\bibfnamefont{Q.}~\bibnamefont{Huang}},
  \bibinfo{author}{\bibfnamefont{E.}~\bibnamefont{Bourret-Courchesne}},
  \bibinfo{author}{\bibfnamefont{G.}~\bibnamefont{Simutis}},
  \bibinfo{author}{\bibfnamefont{R.}~\bibnamefont{Khasanov}},
  \bibinfo{author}{\bibfnamefont{D.~X.} \bibnamefont{Yao}},
  \bibnamefont{et~al.}, \bibinfo{journal}{Physical Review B}
  \textbf{\bibinfo{volume}{98}}, \bibinfo{pages}{18} (\bibinfo{year}{2018}),
  ISSN \bibinfo{issn}{24699969}, \eprint{1807.10703}.

\bibitem[{\citenamefont{Materne et~al.}(2019)\citenamefont{Materne, Bi, Zhao,
  Hu, Amig{\'{o}}, Seiro, Aswartham, B{\"{u}}chner, and Alp}}]{Materne2019}
\bibinfo{author}{\bibfnamefont{P.}~\bibnamefont{Materne}},
  \bibinfo{author}{\bibfnamefont{W.}~\bibnamefont{Bi}},
  \bibinfo{author}{\bibfnamefont{J.}~\bibnamefont{Zhao}},
  \bibinfo{author}{\bibfnamefont{M.~Y.} \bibnamefont{Hu}},
  \bibinfo{author}{\bibfnamefont{M.~L.} \bibnamefont{Amig{\'{o}}}},
  \bibinfo{author}{\bibfnamefont{S.}~\bibnamefont{Seiro}},
  \bibinfo{author}{\bibfnamefont{S.}~\bibnamefont{Aswartham}},
  \bibinfo{author}{\bibfnamefont{B.}~\bibnamefont{B{\"{u}}chner}},
  \bibnamefont{and} \bibinfo{author}{\bibfnamefont{E.~E.} \bibnamefont{Alp}},
  \bibinfo{journal}{Physical Review B} \textbf{\bibinfo{volume}{99}},
  \bibinfo{pages}{020505(R)} (\bibinfo{year}{2019}), ISSN
  \bibinfo{issn}{2469-9950}.

\bibitem[{\citenamefont{Saparov et~al.}(2011)\citenamefont{Saparov, Calder,
  Sipos, Cao, Chi, Singh, Christianson, Lumsden, and Sefat}}]{saparov2011}
\bibinfo{author}{\bibfnamefont{B.}~\bibnamefont{Saparov}},
  \bibinfo{author}{\bibfnamefont{S.}~\bibnamefont{Calder}},
  \bibinfo{author}{\bibfnamefont{B.}~\bibnamefont{Sipos}},
  \bibinfo{author}{\bibfnamefont{H.}~\bibnamefont{Cao}},
  \bibinfo{author}{\bibfnamefont{S.}~\bibnamefont{Chi}},
  \bibinfo{author}{\bibfnamefont{D.~J.} \bibnamefont{Singh}},
  \bibinfo{author}{\bibfnamefont{A.~D.} \bibnamefont{Christianson}},
  \bibinfo{author}{\bibfnamefont{M.~D.} \bibnamefont{Lumsden}},
  \bibnamefont{and} \bibinfo{author}{\bibfnamefont{A.~S.} \bibnamefont{Sefat}},
  \bibinfo{journal}{Phys. Rev. B} \textbf{\bibinfo{volume}{84}},
  \bibinfo{pages}{245132} (\bibinfo{year}{2011}).

\bibitem[{\citenamefont{Bull et~al.}(2016)\citenamefont{Bull, Funnell, Tucker,
  Hull, Francis, and Marshall}}]{bullpearl}
\bibinfo{author}{\bibfnamefont{C.~L.} \bibnamefont{Bull}},
  \bibinfo{author}{\bibfnamefont{N.~P.} \bibnamefont{Funnell}},
  \bibinfo{author}{\bibfnamefont{M.~G.} \bibnamefont{Tucker}},
  \bibinfo{author}{\bibfnamefont{S.}~\bibnamefont{Hull}},
  \bibinfo{author}{\bibfnamefont{D.~J.} \bibnamefont{Francis}},
  \bibnamefont{and} \bibinfo{author}{\bibfnamefont{W.~G.}
  \bibnamefont{Marshall}}, \bibinfo{journal}{High Pressure Research}
  \textbf{\bibinfo{volume}{36}}, \bibinfo{pages}{493} (\bibinfo{year}{2016}).

\bibitem[{\citenamefont{Khasanov et~al.}(2016)\citenamefont{Khasanov, Guguchia,
  Maisuradze, Andreica, Elender, Raselli, Shermadini, Goko, Knecht, Morenzoni
  et~al.}}]{khasanovgpd}
\bibinfo{author}{\bibfnamefont{R.}~\bibnamefont{Khasanov}},
  \bibinfo{author}{\bibfnamefont{Z.}~\bibnamefont{Guguchia}},
  \bibinfo{author}{\bibfnamefont{A.}~\bibnamefont{Maisuradze}},
  \bibinfo{author}{\bibfnamefont{D.}~\bibnamefont{Andreica}},
  \bibinfo{author}{\bibfnamefont{M.}~\bibnamefont{Elender}},
  \bibinfo{author}{\bibfnamefont{A.}~\bibnamefont{Raselli}},
  \bibinfo{author}{\bibfnamefont{Z.}~\bibnamefont{Shermadini}},
  \bibinfo{author}{\bibfnamefont{T.}~\bibnamefont{Goko}},
  \bibinfo{author}{\bibfnamefont{F.}~\bibnamefont{Knecht}},
  \bibinfo{author}{\bibfnamefont{E.}~\bibnamefont{Morenzoni}},
  \bibnamefont{et~al.}, \bibinfo{journal}{High Pressure Research}
  \textbf{\bibinfo{volume}{36}}, \bibinfo{pages}{140} (\bibinfo{year}{2016}).

\bibitem[{\citenamefont{Rodriguez-Carvajal}(1990)}]{fullprof}
\bibinfo{author}{\bibfnamefont{J.}~\bibnamefont{Rodriguez-Carvajal}}, in
  \emph{\bibinfo{booktitle}{Satellite meeting on powder diffraction of the XV
  congress of the IUCr}} (\bibinfo{organization}{Toulouse, France:[sn]},
  \bibinfo{year}{1990}), vol. \bibinfo{volume}{127}.

\bibitem[{\citenamefont{Svitlyk et~al.}(2019)\citenamefont{Svitlyk, Garbarino,
  Rosa, Pomjakushina, Krzton-Maziopa, Conder, Nunez-Regueiro, and
  Mezouar}}]{Svitlyk_2019}
\bibinfo{author}{\bibfnamefont{V.}~\bibnamefont{Svitlyk}},
  \bibinfo{author}{\bibfnamefont{G.}~\bibnamefont{Garbarino}},
  \bibinfo{author}{\bibfnamefont{A.~D.} \bibnamefont{Rosa}},
  \bibinfo{author}{\bibfnamefont{E.}~\bibnamefont{Pomjakushina}},
  \bibinfo{author}{\bibfnamefont{A.}~\bibnamefont{Krzton-Maziopa}},
  \bibinfo{author}{\bibfnamefont{K.}~\bibnamefont{Conder}},
  \bibinfo{author}{\bibfnamefont{M.}~\bibnamefont{Nunez-Regueiro}},
  \bibnamefont{and} \bibinfo{author}{\bibfnamefont{M.}~\bibnamefont{Mezouar}},
  \bibinfo{journal}{Journal of Physics: Condensed Matter}
  \textbf{\bibinfo{volume}{31}}, \bibinfo{pages}{085401}
  (\bibinfo{year}{2019}).

\bibitem[{\citenamefont{Caron et~al.}(2012)\citenamefont{Caron, Neilson,
  Miller, Arpino, Llobet, and McQueen}}]{Caron2012}
\bibinfo{author}{\bibfnamefont{J.~M.} \bibnamefont{Caron}},
  \bibinfo{author}{\bibfnamefont{J.~R.} \bibnamefont{Neilson}},
  \bibinfo{author}{\bibfnamefont{D.~C.} \bibnamefont{Miller}},
  \bibinfo{author}{\bibfnamefont{K.}~\bibnamefont{Arpino}},
  \bibinfo{author}{\bibfnamefont{A.}~\bibnamefont{Llobet}}, \bibnamefont{and}
  \bibinfo{author}{\bibfnamefont{T.~M.} \bibnamefont{McQueen}},
  \bibinfo{journal}{Phys. Rev. B} \textbf{\bibinfo{volume}{85}},
  \bibinfo{pages}{180405(R)} (\bibinfo{year}{2012}).

\bibitem[{\citenamefont{Suter and Wojek}(2012)}]{suter2012}
\bibinfo{author}{\bibfnamefont{A.}~\bibnamefont{Suter}} \bibnamefont{and}
  \bibinfo{author}{\bibfnamefont{B.}~\bibnamefont{Wojek}},
  \bibinfo{journal}{Physics Procedia} \textbf{\bibinfo{volume}{30}},
  \bibinfo{pages}{69} (\bibinfo{year}{2012}).

\bibitem[{\citenamefont{Kobayashi et~al.}(2018)\citenamefont{Kobayashi, Maki,
  Murakami, Hirata, Ohgushi, and ichi Yamaura}}]{Kobayashi_2018}
\bibinfo{author}{\bibfnamefont{K.}~\bibnamefont{Kobayashi}},
  \bibinfo{author}{\bibfnamefont{S.}~\bibnamefont{Maki}},
  \bibinfo{author}{\bibfnamefont{Y.}~\bibnamefont{Murakami}},
  \bibinfo{author}{\bibfnamefont{Y.}~\bibnamefont{Hirata}},
  \bibinfo{author}{\bibfnamefont{K.}~\bibnamefont{Ohgushi}}, \bibnamefont{and}
  \bibinfo{author}{\bibfnamefont{J.}~\bibnamefont{ichi Yamaura}},
  \bibinfo{journal}{Superconductor Science and Technology}
  \textbf{\bibinfo{volume}{31}}, \bibinfo{pages}{105002}
  (\bibinfo{year}{2018}).

\bibitem[{\citenamefont{Fei et~al.}(1995)\citenamefont{Fei, Prewitt, Mao, and
  Bertka}}]{Fei1892}
\bibinfo{author}{\bibfnamefont{Y.}~\bibnamefont{Fei}},
  \bibinfo{author}{\bibfnamefont{C.~T.} \bibnamefont{Prewitt}},
  \bibinfo{author}{\bibfnamefont{H.-k.} \bibnamefont{Mao}}, \bibnamefont{and}
  \bibinfo{author}{\bibfnamefont{C.~M.} \bibnamefont{Bertka}},
  \bibinfo{journal}{Science} \textbf{\bibinfo{volume}{268}},
  \bibinfo{pages}{1892} (\bibinfo{year}{1995}), ISSN \bibinfo{issn}{0036-8075}.

\bibitem[{\citenamefont{Rueff et~al.}(1999)\citenamefont{Rueff, Kao, Struzhkin,
  Badro, Shu, Hemley, and Mao}}]{rueff1999}
\bibinfo{author}{\bibfnamefont{J.-P.} \bibnamefont{Rueff}},
  \bibinfo{author}{\bibfnamefont{C.-C.} \bibnamefont{Kao}},
  \bibinfo{author}{\bibfnamefont{V.~V.} \bibnamefont{Struzhkin}},
  \bibinfo{author}{\bibfnamefont{J.}~\bibnamefont{Badro}},
  \bibinfo{author}{\bibfnamefont{J.}~\bibnamefont{Shu}},
  \bibinfo{author}{\bibfnamefont{R.~J.} \bibnamefont{Hemley}},
  \bibnamefont{and} \bibinfo{author}{\bibfnamefont{H.~K.} \bibnamefont{Mao}},
  \bibinfo{journal}{Phys. Rev. Lett.} \textbf{\bibinfo{volume}{82}},
  \bibinfo{pages}{3284} (\bibinfo{year}{1999}).

\bibitem[{\citenamefont{Lebert et~al.}(2018)\citenamefont{Lebert, Bal\'edent,
  Toulemonde, Ablett, and Rueff}}]{lebert2018}
\bibinfo{author}{\bibfnamefont{B.~W.} \bibnamefont{Lebert}},
  \bibinfo{author}{\bibfnamefont{V.}~\bibnamefont{Bal\'edent}},
  \bibinfo{author}{\bibfnamefont{P.}~\bibnamefont{Toulemonde}},
  \bibinfo{author}{\bibfnamefont{J.~M.} \bibnamefont{Ablett}},
  \bibnamefont{and} \bibinfo{author}{\bibfnamefont{J.-P.} \bibnamefont{Rueff}},
  \bibinfo{journal}{Phys. Rev. B} \textbf{\bibinfo{volume}{97}},
  \bibinfo{pages}{180503(R)} (\bibinfo{year}{2018}).

\bibitem[{\citenamefont{Fernandes et~al.}(2014)\citenamefont{Fernandes,
  Chubukov, and Schmalian}}]{Fernandes2014}
\bibinfo{author}{\bibfnamefont{R.~M.} \bibnamefont{Fernandes}},
  \bibinfo{author}{\bibfnamefont{A.~V.} \bibnamefont{Chubukov}},
  \bibnamefont{and}
  \bibinfo{author}{\bibfnamefont{J.}~\bibnamefont{Schmalian}},
  \bibinfo{journal}{Nature Physics} \textbf{\bibinfo{volume}{10}},
  \bibinfo{pages}{97} (\bibinfo{year}{2014}), ISSN \bibinfo{issn}{1745-2473}.

\bibitem[{\citenamefont{Zhang et~al.}(2018)\citenamefont{Zhang, Lin, Zhang,
  Dagotto, and Dong}}]{Zhang2018}
\bibinfo{author}{\bibfnamefont{Y.}~\bibnamefont{Zhang}},
  \bibinfo{author}{\bibfnamefont{L.-F.} \bibnamefont{Lin}},
  \bibinfo{author}{\bibfnamefont{J.-J.} \bibnamefont{Zhang}},
  \bibinfo{author}{\bibfnamefont{E.}~\bibnamefont{Dagotto}}, \bibnamefont{and}
  \bibinfo{author}{\bibfnamefont{S.}~\bibnamefont{Dong}},
  \bibinfo{journal}{Phys. Rev. B} \textbf{\bibinfo{volume}{97}},
  \bibinfo{pages}{045119} (\bibinfo{year}{2018}).

\bibitem[{\citenamefont{Wu et~al.}(2018)\citenamefont{Wu, Bull, Forrest, Yin,
  Frandsen, and Birgeneau}}]{isisnm}
\bibinfo{author}{\bibfnamefont{S.}~\bibnamefont{Wu}},
  \bibinfo{author}{\bibfnamefont{C.~L.} \bibnamefont{Bull}},
  \bibinfo{author}{\bibfnamefont{T.~R.} \bibnamefont{Forrest}},
  \bibinfo{author}{\bibfnamefont{J.}~\bibnamefont{Yin}},
  \bibinfo{author}{\bibfnamefont{B.}~\bibnamefont{Frandsen}}, \bibnamefont{and}
  \bibinfo{author}{\bibfnamefont{R.~J.} \bibnamefont{Birgeneau}},
  \bibinfo{journal}{STFC ISIS Neutron and Muon Source}  (\bibinfo{year}{2018}),
  \urlprefix\url{https://doi.org/10.5286/ISIS.E.RB1810189}.

\end{thebibliography}
\end{document}



\title{Supplementary Material for `Robust block magnetism in spin ladder BaFe$_2$Se$_3$ under hydrostatic pressure'  }

\author{Shan Wu$^{1*}$}
\author{Junjie Yin$^2$}
\author{Thomas Smart$^3$}
\author{Arani Acharya$^1$}
\author{Craig L Bull$^4$}
\author{Nicholas P Funnell$^4$}
\author{Thomas R Forrest$^5$}
\author{Gediminas  Simutis$^6$}
\author{Rustem Khasanov$^6$}
\author{Sylvia K. Lewin$^1$}
\author{Meng Wang$^2$}
\author{Benjamin A. Frandsen$^7$}
\author{Raymond Jeanloz$^3$}
\author{Robert J. Birgeneau$^1$}

\date{\today}
\begin{abstract}
In support of the main text, the supplementary material includes the following information: (1) detailed experimental set-ups for neutron diffraction, X-ray diffraction and muon spin relaxation ($\mu SR$) measurements under pressure; (2) representative neutron diffraction patterns in the transverse mode under pressure; (3) pressure-dependent x-ray diffraction patterns at room temperature; (4) Rietveld refinement of neutron powder diffraction at $T$ = 1.5 K on the HB3A, HFIR; (5) Full width at half maximum of the magnetic peak $Q_{m1} = (0.5,0.5,0.5)$ versus temperature for various pressure; and (6) Calculations of diffraction pattern with different spin models in the block-type state.
\end{abstract}

\maketitle

Neutron diffraction measurements under pressure employed the Pearl diffractometer with a Paris-Edinburgh (PE) press at the ISIS Pulsed neutron and Muon source, UK \cite{bullpearl}. A mixture of BaFe$_2$Se$_3$ (milligram-scale quantity) and lead powder was pressed in a single-toroidal zirconia-toughened alumina anvil with Ti-Zr gaskets for pressures below 6 GPa and a sintered diamond anvil above 6 GPa. Hydrostatic pressure was provided with a pressure-transmitting fluid of pre-deuterated methanol/ethanol mixture in a 4:1 volume ratio. There are two detector modes: longitudinal and transverse. The longitudinal mode, with three detectors aligned along the beam path, allows access to wave vector transfer Q down to 0.6 $\rm \AA^{-1}$. The transverse mode involves nine detectors located perpendicular to the beam path, and covers the range $1.5 \rm \AA^{-1}< Q < 5 \rm \AA^{-1}$. The base temperature was reached by immersing the PE press in liquid nitrogen with an uncertainty of 2-3 K.   
The lowest accessible temperature,120 K, could be reached by immersing the PE press in liquid nitrogen in the cryostat tank. The pressure was calibrated with the lattice constant of the lead with an uncertainty in the measured pressure of 0.3 GPa. The significantly reduced intensity in the longitudinal mode necessitates counting for several hours to achieve adequate statistics for each diffraction pattern. The intensities have been normalized to  the incident beam monitor for one hour and corrected for the detector efficiency. 
Muon spin relaxation ($\mu SR$) spectroscopy was performed on samples under pressure using the general purpose decay-channel (GPD) instrument at the Paul Scherrer Institute, Switzerland \cite{khasanovgpd}. The body of the pressure cell used a MP35N(Ni-Co-Cr-Mo) alloy. Daphne oil was used as pressure-transmitting fluid. The pressure was calibrated with a superconducting indium plate immersed in the oil with an uncertainty in the measured pressure less than 0.1 GPa. 
A gas-flow cryostat was used for temperature control. 
Room temperature pressure-dependent X-ray diffraction was carried out at the Lawrence Berkeley National Laboratory Advanced Light Source beam line 12.2.2. We loaded the mixture of the sample and ruby powder into a Merrill-Basset type diamond anvil cell with a steel gasket and 4:1 methanol/ethanol pressure-transmitting fluid. The pressure was calibrated by the ruby fluorescence R1-line with pressure accuracy of 0.1 GPa.

\begin{figure}[h!]
\includegraphics[width=1\columnwidth,clip,angle =0]{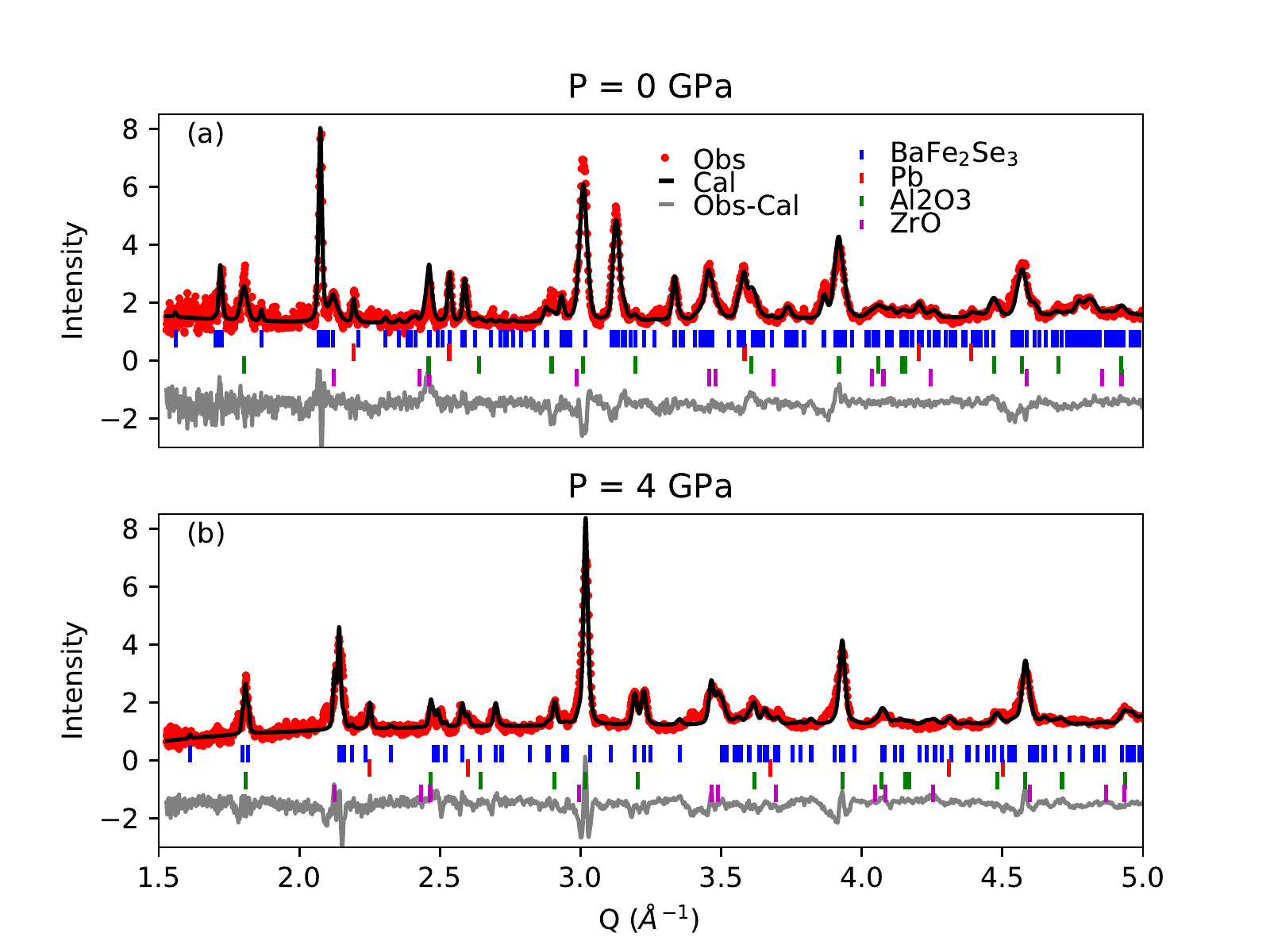}
\caption{\label{pearl} Room temperature neutron powder diffraction data at $P$ = 0 and 4 GPa  collected  in the transverse mode (Pearl diffractometer, ISIS) and their corresponding Rietveld refinements. }
\end{figure}

Representative neutron powder diffraction data collected in the transverse mode are shown in Fig. \ref{pearl}. $P$ = 0 GPa and 4 GPa data are fitted with the space group $Pnma$  and $Cmcm$ respectively, consistent with x-ray diffraction measurements under pressure displayed in the main text Fig. 1 (b). The fitting patterns include the phases from the lead powder used for the pressure calibration and the pressure cell polymer of Al$_2$O$_3$ and ZrO. 

\begin{figure}
\includegraphics[width=1\columnwidth,clip,angle =0]{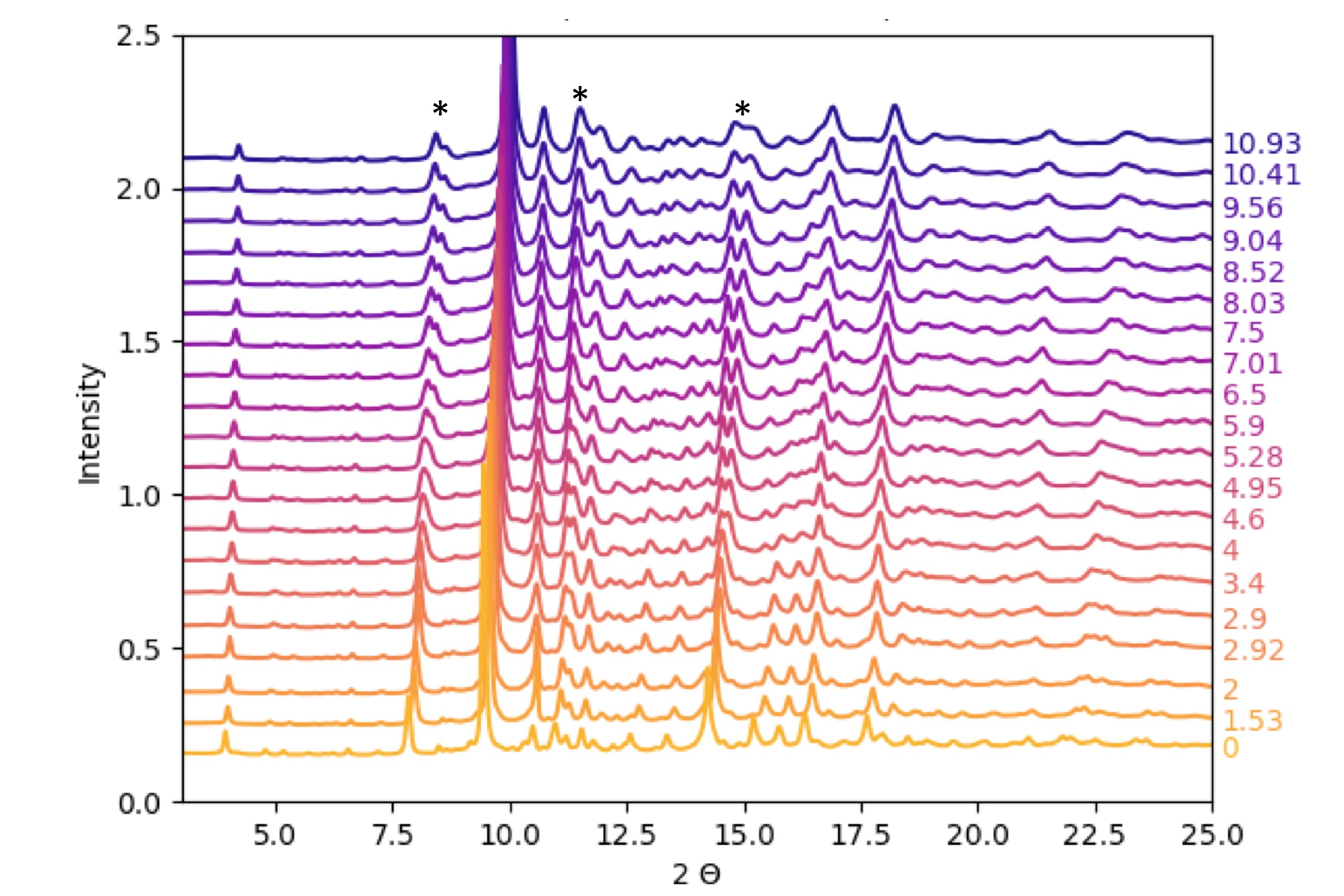}
\caption{\label{pdep} Pressure-dependent X-ray powder diffraction patterns from  $P$ = 10.93 GPa to ambient pressure (decompression data). The values shown on the right of the plot are in GPa. Asterisks indicate peaks showing
evidence of a crystal-structural transformation.}
\end{figure}

Pressure-dependent X-ray diffraction measurements show evidence of a structural transition (Fig. \ref{pdep}). The higher crystallographic symmetry of the $Cmcm$ phase leads to a reduced number of  Bragg peak positions. For example, at the position of $2\theta \sim 7.6^{\circ}$ the peak is indexed as the superposition of  Bragg peak (211), (301) and (202) in the $Pnma$ phase, while the (211) peak is forbidden in the $Cmcm$ phase (Rietveld refinements are shown in the Fig. 1(b) of the main text). The extracted lattice parameters from compressed and decompressed data display a change of slope marked by the results of linear fits in the Fig. 4 (d) of the main text. The structural phase transition has thus been validated by both neutron and x-ray diffraction measurements.

\begin{figure}
\includegraphics[width=1\columnwidth,clip,angle =0]{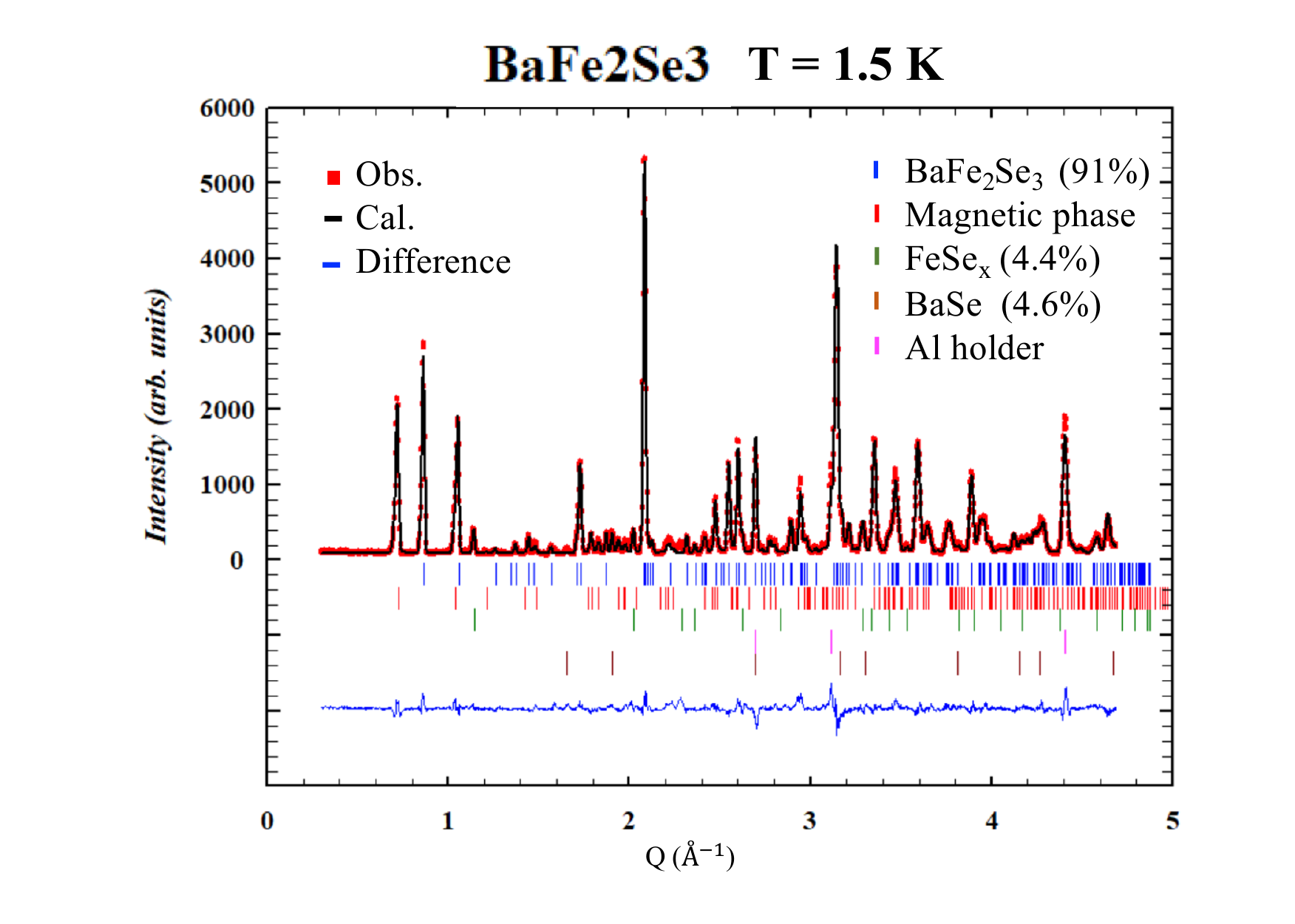}
\caption{\label{hfir} Neutron powder diffraction data at $T$ = 1.5 K collected on the HB3A, HFIR and its corresponding Rietveld refinement. }
\end{figure}

A neutron powder diffraction pattern at $T$ = 1.5 K and its corresponding Rietveld refinement are shown in  Fig. \ref{hfir}. The diffraction pattern consists of the contributions from  91\% BaFe$_2$Se$_3$, along with 4.4\% FeSe$_x$ and 4.6\% BaSe impurity phase. Magnetic structure refinement with a block-type spin structure  provides a good fit to the data.

By fitting the leading magnetic peak $Q_{m1} = (0.5,0.5,0.5)$  using a single Gaussian profile, the full width at half maximum (FWHM) is depicted at various temperatures and pressures (Fig. S4). The dashed grey line gives the instrumental $Q$-resolution obtained from a nearby nuclear peak, so all of the peaks at temperatures below $T_N$ are $Q$-resolution limited within present uncertainties. 

Fig. \ref{spin} shows the difference pattern between 120 K (T $<$ T$_N$) and 260 K (T $>$ T$_N$) collected at $P$ = 4 GPa. The wave vector transfers $Q$ at 0.75 $\rm \AA^{-1}$ and 1.1 $\rm \AA^{-1}$ are associated with block-type order and can be indexed as $Q_{m1}$ = ($\frac{1}{2},\frac{1}{2},\frac{1}{2}$) and $Q_{m2}$ = ($\frac{1}{2},\frac{3}{2},\frac{1}{2}$) in the $Cmcm$ phase, respectively. The slight offset for the second $Q$ position could be attributed to poor statistics of the relative weak signal.  The red lines in the three panels display the calculated patterns with the spins oriented along three crystallographic axes. The main discrepancy is the different relative intensities between $ Q_{m1}$ and $Q_{m2}$. The calculated intensities with spins pointing perpendicular to the ladder direction ($c$ axis) agree better with the observed data.   The best fit (statistically)  is associated with spins along the rung direction ($a$ axis), leading to spins laying down in the ladder plane.  This might suggest a spin reorientation from out-of-ladder to within-ladder plane, accompanying the structural phase transition. However, more studies are required to confirm this hypothesis.     

\begin{figure}[h!]
\includegraphics[width=0.8\columnwidth,clip,angle =0]{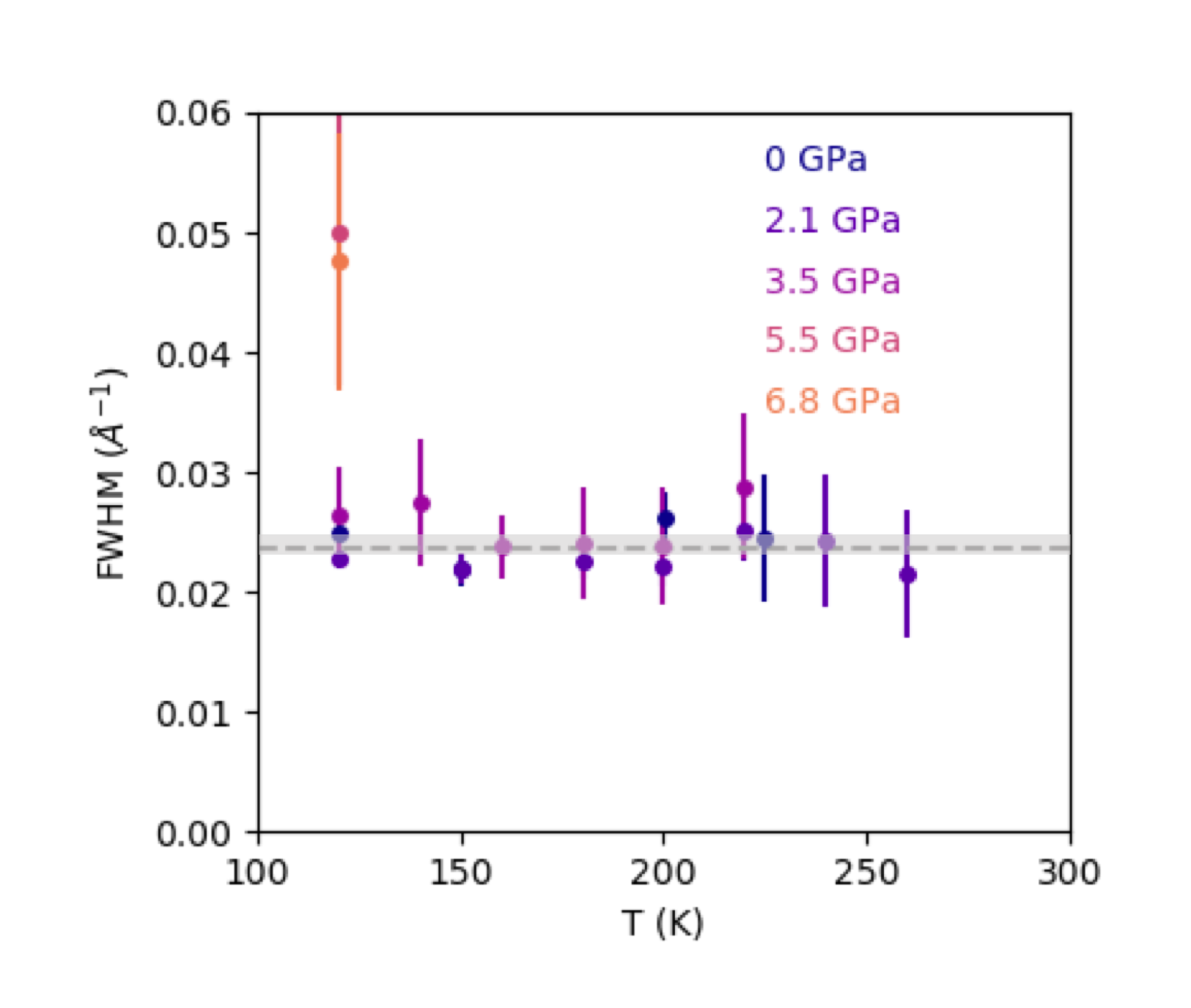}
\caption{\label{wd} Full width at half maximum (FWHM) of the magnetic peak  $Q_{m1} = (0.5,0.5,0.5)$ versus temperature for various pressures. The dashed grey line shows the $Q$-resolution extracted from the nearby (101) Nuclear Bragg peak. }
\end{figure}

\begin{figure}[h!]
\includegraphics[width=0.8\columnwidth,clip,angle =0]{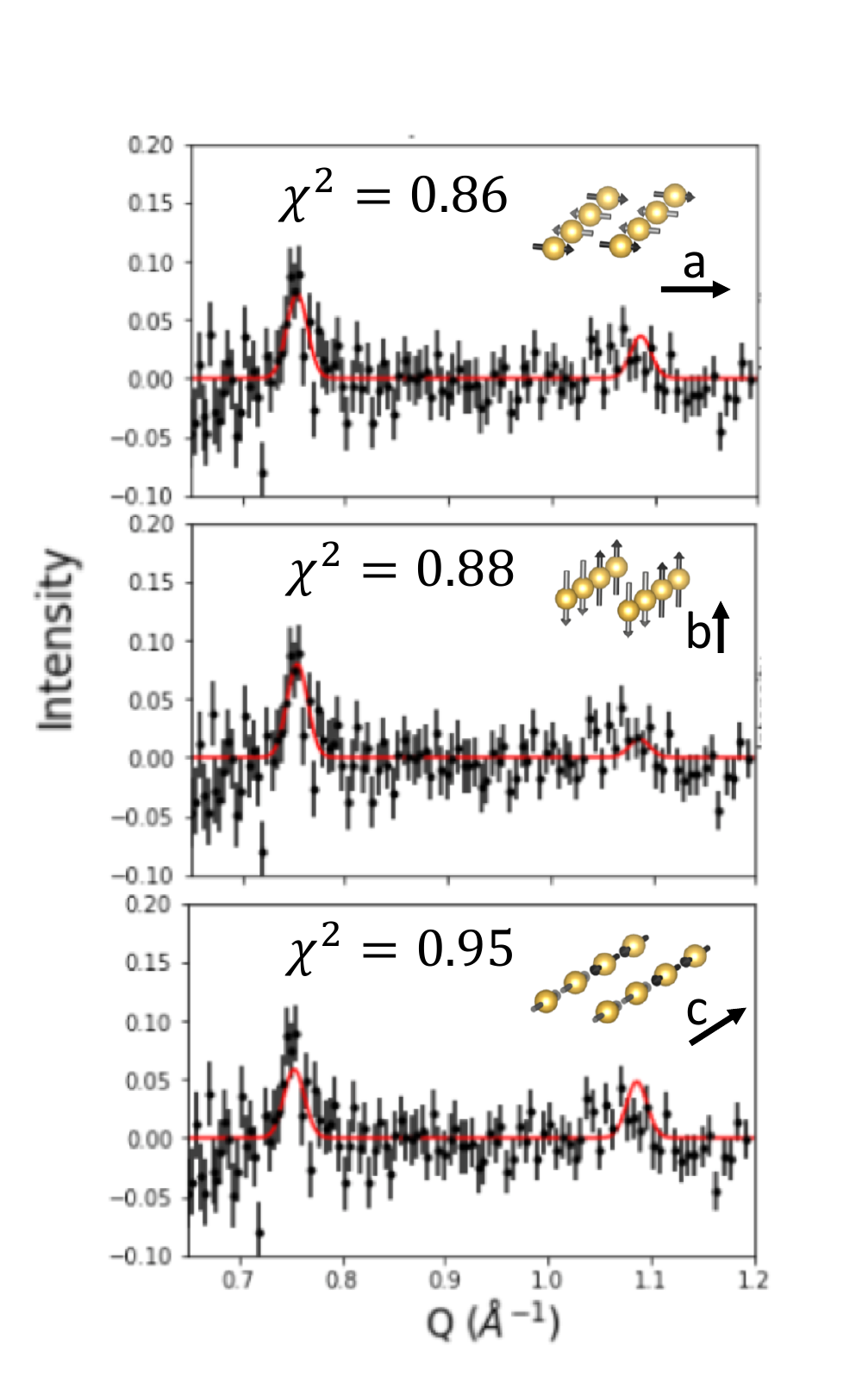}
\caption{\label{spin}  Calculated diffraction pattern with spins along $a$,$b$ and $c$ axes for the block-type order in the $Cmcm$ phase ($P$ = 4 GPa). The black dots give the difference patterns between 120 K ($T <$ $T_N$) and 260 K ($T >$ $T_N$), with reduced $\chi^2$ calculated as $\sum_i \frac{1}{N-1}\frac{(y_{i,o}-y_{i,c})^2}{dy_{i,o}^2}$ where $y_{i,c}$ and $y_{i,o}$ are calculated and observed intensity.}
\end{figure}

We can determine the magnetic volume fraction from muon spin relaxation measurement in a weak transverse field (TF). The asymmetry spectrum with a magnetically ordered volume fraction $f$, the remaining paramagnetic fraction $(1-f)$ and the contribution from the pressure cell are described by the expression:  $P(t) = A(\frac{1}{3}fe^{-\lambda_1t}+(1-f)cos(\gamma_{\mu}Bt)e^{-\lambda_2t})+ P(t)_{NP}$. Here $A$ is the total asymmetry; $B$ is the applied external weak transverse field; $\lambda_1$ and $\lambda_2$ are the longitudinal relaxation rate due to the internal field fluctuating in time and paramagnetic spin fluctuations respectively. $P(t)_{NP}$ is the portion from the pressure cell which follows the revised Gaussian depolarization function  \cite{khasanovgpd}. The magnetic volume fraction $f$ can be extracted from the fitting to the weak TF oscillation asymmetry patterns. The value of $f$ can also be obtained from longitudinal zero field measurements that fit to an exponentially decaying and Kubo-Toyabe depolarization function  \cite{khasanovgpd}. The static and dynamic relaxation rates are interpolated according to the literature \cite{khasanovgpd}.

\bibliography{bibfile}